\documentclass{article}
\usepackage{arxiv}

\usepackage[utf8]{inputenc} 
\usepackage[T1]{fontenc}    
\usepackage{hyperref}       
\usepackage{url}            
\usepackage{booktabs}       
\usepackage{amsfonts}       
\usepackage{nicefrac}       
\usepackage{microtype}      
\usepackage{lipsum}

\usepackage{pgfplots}
\usepackage{graphicx}
\usepackage{tikz}
\pgfplotsset{
  width=0.95\textwidth,
  height=0.2\textheight,
   grid=major,
   major grid style={dotted},
   enlarge y limits={upper,value=0.05},
   }
 
\usepackage{url}
\usepackage{varwidth}
\usetikzlibrary{shapes.geometric, arrows}

\title{An Empirical Study on User Reviews Targeting Mobile Apps' Security \& Privacy}

\author{
 Debjyoti Mukherjee \\
  Department of Computer Science\\
  University of Calgary, Canada \\
  \texttt{debjyoti.mukherje1@ucalgary.ca} \\
   \And
 Alireza Ahmadi \\
  Department of Computer Science\\
  University of Calgary, Canada\\
  \texttt{alireza.ahmadi1@ucalgary.ca} \\
   \AND
   Maryam Vahdat Pour \\
   Schulich School of Engineering \\
   University of Calgary, Canada \\
   \texttt{maryam.vahdatpour@ucalgary.ca} \\
   \And
   Joel Reardon \\
   Department of Computer Science \\
   University of Calgary, Canada \\
   \texttt{joel.reardon@ucalgary.ca} \\
}

\begin{document}

 \tikzstyle{rect} = [rectangle, rounded corners, text centered, draw=black, fill=white!30]
 \tikzstyle{elli} = [ellipse, fill=white!20, minimum height=1cm, text centered, draw=black]
 \tikzstyle{io} = [trapezium, trapezium left angle=70, trapezium right angle=110, text centered, draw=black, fill=white!30]
 
 \tikzstyle{circ} = [circle, fill=white!20, draw=black, text centered, inner sep=0.1cm]

 \tikzstyle{decision} = [diamond, fill=white!20, text width=6em, text centered, inner sep=0pt]
 \tikzstyle{line} = [draw, -latex']
 \tikzstyle{arrow} = [thick,->,>=stealth]

\maketitle

\begin{abstract}
    Application markets provide a communication channel between app developers and their end-users in form of app reviews, which allow users to provide feedback about the apps. Although security and privacy in mobile apps is one of the biggest issues, it is unclear how much people are aware of these or discuss about them in reviews.\\
  In this study, we explore the privacy and security concerns of users using reviews in the Google Play Store. For this we conducted a study by analyzing around 2.2M reviews from the top 539 apps of this Android market. We found that 0.5\% of these reviews are related to the security and privacy concerns of the users. We further investigated these apps by performing dynamic analysis which provided us valuable insights into their actual behaviours. Based on the different perspectives, we categorized the apps and evaluated how the different factors influence the users' perception about the apps. It was evident from the results that the number of permissions that the apps request plays a dominant role in this matter. We also found that sending out the location can affect the users' thoughts about the app. The other factors do not directly affect the privacy and security concerns for the users.
\end{abstract}

\keywords{Natural Language Processing \and Google Play \and App Review \and Mobile Application \and Security \and Privacy \and Dynamic Analysis}

\section{Introduction}

Mobile application have been a part of computers for over a decade and mobile software market is the fastest growing segment in the mobile industry. With the ever-growing popularity of mobile apps, various OS providers and device vendors have launched their own application stores; Google Play and Apple's App Store are the two most popular among them. These markets distribute many apps for end-users to search, download, and purchase applications. Similar to online retail markets, end-user reviews are a key for the success of the apps. Users that have used an app can write reviews---including a 1-to-5-star rating---to express their opinion about an app and help other users to choose among similar apps. 

Reviews can also be used as a direct feedback channel to app developers. Developers can find out feature suggestions, as well as usability issues, crashes, and other types of feedback about their apps. While prior research has focused on providing users with support for choosing less risky apps~\cite{b1, b2} or helping them making informed decisions~\cite{b3, b4, b5}, there is a dearth of research related to this feedback channel. We believe that for apps to improve their security-and-privacy related behavior, feedback should be directed to developers. User reviews would seemingly form such an immediate feedback and rating channel for security and privacy related concerns from users.

This paper performs an empirical study on mobile app reviews for top free mobile applications in Google Play to explore how much the users are concerned about their security-and-privacy while using mobile apps and whether their concerns are justified. In this study, we have answered the following two research questions: \\
\textit{\textbf{RQ 1:} } How much are users concerned about security and privacy while using mobile apps? \\
\textit{\textbf{RQ 2:} } To what extent, users’ judgment matches the actual functionality of mobile apps?

To answer RQ1, we mined the reviews and apps' details from the app store and performed supervised learning to identify the security-and-privacy related reviews. We refer to these reviews as \textit{``Related''} reviews and observed that a considerable number of reviews are related. We also devised a method to identify if the related reviews correspond to a positive or a negative sentiment. Based on this classification, we have successfully labelled the apps.

In order to evaluate RQ2, we performed dynamic analysis to check the actual behaviour of the apps and collected various types of data like Personal Identifiable Information (PII) leaked, the different hosts connected, the type of permissions asked, etc. On the basis of these data we have again categorized the apps. Finally we have compared the categorization of apps based on reviews to new categories based on the dynamic analysis and have successfully answered the second research question. We have also used different statistical measures to evaluate our results.

In summary, the contributions of this work are the following:

\begin{itemize}
    \item We demonstrate a way to judge how much the users of mobile apps are concerned about their privacy-and-security
    \item We have measured the actual behavior of the apps as it relates to privacy and could compare how much of the users' concerns are justified
    \item We have identified some of the key factors that users' depend on while judging any app's behavior related to privacy
    \item We show that reviews can be useful for identifying some types of privacy violation in mobile apps, while it may not be as effective in some other aspects
\end{itemize}

The rest of this paper is structured as follows: Section~\ref{sec_related} provides an overview of the related work, Section~\ref{methodology} depicts our methodology, Section~\ref{dataset} provides overview of the data analyzed, Section~\ref{result} shows the results, Section~\ref{sec_discussion} contains the discussion, and Section~\ref{sec_conclusion} has the conclusion.

\section{Related Work} \label{sec_related}

Android privacy, and in particular application privacy and the role of developers in the mobile ecosystem, have been studied from a variety of perspectives. In this section, we survey related works on app reviews in general and how describe privacy issue and awareness on apps using natural language. We would also take a look at app security evolution.

\subsection{General App Reviews} 
App reviews are the primary channel through which developers receive feedback about their applications.
Prior work by Pagano and Maalej~\cite{b6} found that different apps receive different amount of reviews, and reviews are not easy to automatically analyze given their unstructured forms. 
Chen et al.~\cite{b7} has shown that about one third of the user reviews are informative and focused on automatically identifying useful user reviews for developers. Existing work by Palomba et al.~\cite{b8} also proposed a similar approach to support app developers in classifying feedback useful for app maintenance. Fu et al.~\cite{b9} proposed a tool that analyzes user comments and ratings in mobile app markets. The approach uses regression and Latent Dirichlet Allocation (LDA)~\cite{b10} models to analyze the comments’ topics. In contrast, our study focuses on the connection between app reviews and the application’s security and privacy concerns.

\subsection{Developer Reviews} Past research has successfully mined software artifacts and connected them with the app descriptions regarding security and privacy aspects. Gorla et al.~\cite{b2} proposed an approach to examine whether the applications’ descriptions matches the applications’ behavior. 
It offers a solution to cluster apps by their topics based on their description, and the usage of permission for protected APIs within each cluster. Further, Pandita et al.~\cite{b11} and Qu et al.~\cite{b1} proposed two systems that mine Android application descriptions and then use natural language processing (NLP) to automatically bridge the semantic gap between what applications do and what users expect them to do from their description. Our study, however, focuses on reviews written by users, which do not always follow rigid grammatical structures~\cite{b12,b13,b14}.

Recent works by Gruber et al.~\cite{b15} also focused on mining privacy policy of apps to identify critical discrepancies between developer-described app behavior and permission usage. Further, Sadeh et al.~\cite{b16} focus is on comparing the practices described in privacy policies to the practices performed by smartphone apps covered by those policies. Although these two works are comparing privacy policy of the apps, it is not based on real experience, and it is based on the documents provided by the developers.

Bugiel et al.~\cite{b17} measured the impact of
user reviews on Android app security and privacy. This method first measures the security and privacy relevant reviews (SPR), and then for each app version mentioned in the SPR, they use static code analysis to
extract permission-protected features mentioned in the reviews. However, their study does not show if the mentioned privacy and security leaks is actually leaking in the application or not.

\subsection{App Security Evolution}
 Calciati et al.~\cite{b18} studied how the permissions requested by apps evolve across different app versions. Their results show that many newly requested permissions are in apps evolution.
 Felt et al.~\cite{b19} identified the violation of least-privilege by app developers, which is unfortunately a long-standing problem. Past research has also investigated how users should be confronted with permission requests, most noticeably early studies by Felt et al.~\cite{b20,b21}. More disruptive proposals try to eliminate the explicit role of the user for permission granting as proposed by Roesner et al.~\cite{b5} or the use of machine learning as proposed by Wijesekera et al.~\cite{b4} and Olejnik et al.~\cite{b3}. 
 Most recently, different works pointed out the risks of third party libraries, in particular of advertisement libraries~\cite{b22,b23,b24,b25} and other vulnerable libraries \cite{b26, b27}. However, to the best of our knowledge, we are the first to study how much app details can be connected to the security and privacy related reviews.


\section{Methodology} \label{methodology}
\begin{figure*}[h]

    \begin{center}
        \begin{tikzpicture}[node distance = 1.5cm, auto]
        \node (start) [circ] {START};
        \node (shape9) [io, right of =start, node distance = 3cm]{Apps' Details}; 
        \node (shape1) [rect, above of=shape9, node distance = 2cm]{%
          \begin{varwidth}{6em}
              Scrape Google Play Store
            \end{varwidth}};
        \node (shape3) [io, right of=shape1, node distance = 3.5cm]{Apps' Reviews};

        \node (shape4) [rect, right of=shape3, node distance = 3.5cm]{%
          \begin{varwidth}{2cm}
              Data Pre-Processing
            \end{varwidth}};
        \node (shape6) [rect, right of=shape4, node distance = 3.5cm] {Build Training Set};
        \node (shape8) [rect, below of=shape6, node distance = 1.5cm] {Classifying};
        \node (shape10) [io, left of=shape8, node distance = 4cm] {%
          \begin{varwidth}{5em}
              Classified Reviews RQ1
            \end{varwidth}};
        \node (shape11) [io, left of=shape10, node distance = 3cm]{%
          \begin{varwidth}{5em}
              Labelled Apps
            \end{varwidth}};

        \node (shape12) [io, below of=start,  node distance = 2.5cm]{Apps};
        \node (shape13) [rect, right of=shape12, node distance = 3cm] {Dynamic Analysis};
        \node (shape14) [io, right of=shape13, node distance = 4cm]{%
          \begin{varwidth}{5em}
              PIIs Hosts Encryption ....
            \end{varwidth}};
        \node (shape15) [rect, right of=shape14, node distance = 4.5cm] {Categorized Apps};
        
        \node (shape16) [rect, above of=shape15, node distance = 1.5cm] {%
          \begin{varwidth}{4em}
              Compare Results RQ2
            \end{varwidth}};
        \node (end) [circ, right of=shape16, node distance = 3.5cm] {END};

        \draw [arrow] (start) -- (shape1);
        \draw [arrow] (shape1) -- (shape3);
        \draw [arrow] (shape1) -- (shape9);
        \draw [arrow] (shape3) -- (shape4);
        \draw [arrow] (shape4) -- (shape6);
        \draw [arrow] (shape6) -- (shape8);
        \draw [arrow] (shape8) -- (shape10);
        \draw [arrow] (shape10) -- (shape11);
        \draw [arrow] (start) -- (shape12);
        \draw [arrow] (shape12) -- (shape13);
        \draw [arrow] (shape13) -- (shape14);
        \draw [arrow] (shape14) -- (shape15);
        \draw [arrow] (shape11) |- (shape16);
        \draw [arrow] (shape15) -- (shape16);
        \draw [arrow] (shape16) -- (end);
        \end{tikzpicture}
    \end{center}
    \caption{Process Flow Diagram}
    \label{fig:processflow}
\end{figure*}
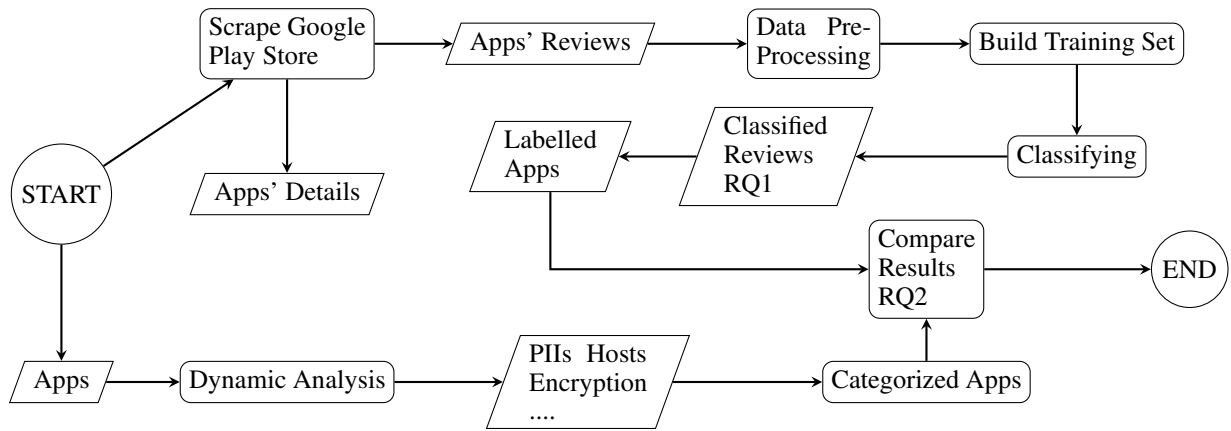

In this section, we discuss the methodologies that we used for our study in order to answer our research questions. To achieve this, we perform two types of activities: identify what the users say about the apps in the reviews, and identify the behaviour of the app. The following sub-sections describe this process in details. The Figure~\ref{fig:processflow} depicts the process flow.

\subsection{Analysing Reviews}
In this section, we discuss the technical details for scraping and classifying the reviews.

\subsubsection{Google Play Scraper} \label{scraper}
 We built a custom web scraper to collect Android applications' details and reviews from Google Play. As previous studies~\cite{b28, b29} have shown that only a small fraction of free applications on Google Play accounts constitute the bulk of the application downloads, we collected the details of the top free apps that are the most popular in Google Play. This resulted in 539 distinct applications. We also used another scraper to mine reviews of these apps. 
 
 We scraped reviews that were written in English language only. After we had scrapped all the reviews for the apps, we pre-processed the reviews. Since user reviews are often written on smart phones, they tend to be short and usually contain grammatical mistakes or typos \cite{b12,b13,b14}. For this sake, we did the following steps:

\begin{itemize}
    \item Removed links, unwanted characters, non-ASCII characters, special characters, etc.
    \item Removed english stop words like ``a'', ``the'', ``from'', ``is'', etc.
    \item Lemmatized the text.
\end{itemize}

The output of the above text pre-processing resulted in the final data that we used for analysis.

\subsubsection{Classifying reviews} \label{classifier}
In order to classify the reviews, we initially needed to build the training set. This is achieved by manually labeling the reviews. Our first step was to find reviews that can be potentially related to security and privacy. We decided to search into reviews using a list of keywords. We initially performed a literature review to identify some related keywords that users may use in their reviews when they describe any security or privacy related aspect. We searched based on these keywords and also manually examined some other reviews to identify other keywords. After some iterations, we could built a list of words that can be used to retrieve related reviews. Table~\ref{tab:keywords} presents this list.

\begin{table}[t]
\centering
\normalsize
\caption{List of Keywords}
\label{tab:keywords}
\begin{center}
 \begin{tabular}{|p{2cm}|p{2cm}|p{2cm}|}
 \hline
 \multicolumn{3}{|c|}{\textbf{Keywords}}\\
  \hline
 Privacy & Security & Safe\\
 Secure & Permission & Identity\\
 Personal & Virus & Malware\\
 Malicious & Access &   Fishy\\
 Phishing & Fishing& Stealth\\
 Steal & Thief & Creepy \\
 \hline
\end{tabular}
\end{center}
\end{table}

We counted the number of occurrences of the keywords in each review. We found that the maximum occurrences is 5 keywords in a single review. Eventually, for the training set, we picked all the reviews with 2 or more keywords (totally 2122 reviews). We also added 2000 reviews, randomly selected from the ones with 1 keyword. Finally, we added 1878 reviews without any keyword to create a fair representative of all the reviews. So our final training set contained a total of 6000 reviews. 

The next task was to manually annotate these reviews as either \textit{Related} or \textit{Non-Related}. For labeling, we considered a review as \textit{Related} if the review matched any of the below criterion:
\begin{itemize}
    \item Concerns about their personal information stolen, illegally accessed, or shared with third parties without permission.
    \item Concerns about their Password or User-name(identity) safety.
    \item Concerns about hidden background activities of the application.
    \item Concerns about unrelated taken permissions.
    \item Concerns about application security/privacy in general.
\end{itemize}

Three of our team members were assigned the task of manually coding these reviews. Each of them individually coded all the 6000 reviews as related or non-related. Once the coding was completed by all the members, the results were matched. For most of the reviews, there was unanimity amongst all the members. For the remaining mismatched labels, we individually discussed to agree on a single label. After completing this process, our training set contained 936 reviews, labelled as \textit{Related} while the remaining 5064 were labelled as \textit{Non-Related}. In order to balance the training set, we performed SMOTE, a well-known over-balancing technique in practice~\cite{b30}.

With our training set, we are now ready to classify the reviews. We needed to evaluate the efficiency of four classifiers, ``Naive Bayes'', ``K-Nearest Neighbour'', ``Single-Layer averaged Perceptron'', and ``Support Vector Machine (SVM)'', known for dealing with text data. To evaluate the efficiency, we used 10-fold cross validation, one of the highly recommended methods for validation~\cite{b31}. We identified that SVM achieved the best results as compared to the other classifiers. It achieved high accuracy, precision and recall; so we selected SVM as the ideal classifier for our analysis.

\subsection{Dynamic analysis}\label{dynamic_analysis}

In this section, we use established dynamic analysis methods~\cite{reyes2018won}
to measure the actual behaviour of the apps as it relates to privacy. In particular, we use an instrumented version of the Android Nougat operating system, which we deployed on actual Nexus 5X phones. This instrumentation monitored all network traffic, including TLS-secured traffic. Our instrumentation can attribute specific network transmissions to the responsible application and records where on the Internet the data was sent. We search through this network traffic to find the presence of types of PII including location data and persistent identifiers.

We tested each of the apps by installing it, granting all runtime permissions, and then using a UI fuzzer to automatically interact with the app for a period of ten minutes. After this time the app is uninstalled and its network transmissions are saved for processing. This processing involves applying a suite of decoders, such as base64 and gzip, to reveal the raw data being transmitted. This also includes a number of deobfuscation methods based on ad-hoc obfuscation methods that we have seen third party libraries use in practice.

\subsubsection{PII Types}

In the network traffic generated by our experiment, we search for the presence of two types of PII: data used to geolocate the user and persistent identifiers for tracking. Location data is either the GPS coordinates, or the SSID and MAC address of the connected WiFi router, which is a well-known surrogate for location. For persistent identifiers, we divide them into two categories: resetable, which consists of the resetable advertising ID (AAID), and non-resetable, which consists of all other identifiers, including the android ID, IMEI, network MAC address, and serial number. 

We separate these types of tracking identifiers because Google recommends developers only use the advertising ID and no other identifier for advertising purposes, and further recommends to avoid bridging resets of the advertising ID by linking it with other identifiers. As such, sending the AAID alone is reasonable when compared to combining it with other non-resetable trackers like the IMEI and the MAC address. 

Based on this information, we categorized each app into one of the following categories:

\begin{itemize}
    \item Good: If the app does not leak any PII type or only the AAID
    \item SingleTracker: If the app only leaks a single tracker PII type and no other PII types
    \item MultiTracker: If the app leaks multiple tracker types of PIIs and no other PII types
    \item SingleLocInfo: If the app leaks only a single Location Info type of PII and no other PII types
    \item MultiLocInfo: If the app leaks multiple Location Info types of PIIS and no other PII type
    \item AAID \& Tracker: If the app leaks AAID and at least one type of Tracker PII
    \item AAID \& LocInfo: If the app leaks AAID and at least one type of Location Info PII
    \item Tracker \& LocInfo: If the app leaks both Tracker and Location Info types of PIIS but not AAID, and
    \item All: If the app leaks all three types of PIIS; i.e AAID, Location Info and Tracker
\end{itemize}

This list does not mean to order them in terms of invasiveness. That is, some may consider location worse than trackers and others feel the opposite. Nevertheless, there is an implicit order based on the subset relation, where sending the AAID is better for privacy than sending the AAID \emph{and} location.

\subsubsection{Domains contacted}

Another metric we used to measure app privacy is the number of different domains to which PII was sent. That is, one app may include a single advertising SDK  while another includes half a dozen so as to maximize revenue. This metric is not perfect, as a half dozen ``good'' SDKs may still be preferable to one invasive one. Nevertheless, the number of places on the Internet that are collecting PII from users devices does indicate how the app developer that includes these SDKs feels about user privacy.

In order to evaluate this, we categorized each app to one of the following groups:

\begin{itemize}
    \item Level 0: 0--1 domains received PII
    \item Level 1: 2 domains received PII
    \item Level 2: 3--6 domains received PII
    \item Level 3: more than 6 domains received PII
\end{itemize}

\subsubsection{PIIs leaked to each domain}

The total number of hosts communicated with by an app may not always reveal the actual nature of invasion. For example, while some apps may be leaking the same PII to a large number of domains, there can be some apps which leak a large number of PIIs to a small number of domains. The impact of these two behaviours would naturally be different. So in addition to the total number of PIIs leaked and the total number of domains contacted, we also elicited the number of PIIs sent by the app to individual domains. We calculated the maximum number of PIIs sent by any app to a single domain.

\subsubsection{Number of permissions asked by the app}

We initially extracted the different types of permissions that an app can ask for and labelled each of them as either ``Normal'' or ``Dangerous'' based on the protection level set in \cite{developer_android}. Dangerorus permissions protect sensitive user data and sensors, like camera, location, and contact lists. As of Android Marshmallow, apps show a runtime dialog asking about the permission at the time it is first used.  
Once we have this data, we evaluated the total number of dangerous and normal permissions that the app needs. In order to judge if the app is encroaching into the users' privacy and security, we are only concerned about the dangerous permissions; also the users will be aware of the dangerous permissions only as the app would specifically request for those permissions. So, for analyzing if the users are concerned about their privacy and security, we consider only the dangerous permissions in our study. 

\subsection{Statistical approaches} \label{stats}

In order to answer RQ2, we needed to calculate the correlation between different categories obtained from methods described above. As we have different types of data (Numerical and Categorical), we used the following statistical methods.
\begin{enumerate}
    \item \textbf{Cramér's V}: A measure of association between two categorical variables. This measure assumes a symmetrical approach; i.e., the correlation between the two variables does not depend on the order of the variables.
    \item \textbf{Theil’s U}: Also known as ``Uncertainty Coefficient'', a measure of categorical association. This measure is used to calculate the correlation between 2 categorical variables when they assume an asymmetric approach.
    \item \textbf{Correlation Ratio}: A measure to calculate the correlation between 2 variables that have mixed data types; i.e., one of the variables is categorical type and the other is of the numerical data type. 
    \item \textbf{Non-parametric Statistical Significance Test}: In order to ascertain if the correlation measures obtained using the above methods are meaningful, we used two well renowned non-parametric statistical significance tests, \textbf{``Mann-Whitney U''} and \textbf{``Kruskal-Wallis H''}. Hereafter, we denote these two test as ``MWU'' and ``KWH'' respectively. We performed the significance test using the data from dynamic analysis on the categories of the apps from reviews' classifier.
\end{enumerate}

\section{Dataset} \label{dataset}
To make a dataset of user reviews about mobile apps, we targeted the ``top free'' list of Google Play, the well-known Android app market powered by Google. The apps in this list are pretty popular and can potentially provide us more reviews to be assessed. This list typically contains 540 apps, but our dataset has 539 apps, as one of the apps was removed during the process of scraping. For each app, we scraped details such as category of the app, score, developer, title, number of reviews, number of installations, chosen by app store editors or not, description, content rating, and number of 1 to 5 stars ratings.

We also scraped the reviews of each app. Due to restrictions in Google API, we have access to the latest 4480 reviews of each app. This means that for less popular apps with less than 4480 reviews, we scraped all the reviews from the time it’s been published. At the same time for some more famous apps, the limit of 4480 provide us reviews from last few months. For each review, we scraped details such as review text, review date, and current rating. For 539 top free apps in Google Play, we were able to scrape 2,186,093 reviews. By looking into the dataset, here we have some statistics:

\begin{enumerate}
    \item App Reviews Rating: For each app, we have average of 2,090,749 ratings. Facebook, WhatsApp, Instagram, Messenger and Clash of Clans have more than 85M, 84M, 78M, 65M and 48M ratings respectively.
    \item App Developers: These 539 apps are developed by 409 different developers, and 57 developers developed more than 1 app. Google with 29 developed apps, Voodoo with 13, Microsoft with 7, and Samsung and Amazon with 6 developed apps are top in the list.
    \item App Category: Apps are from 32 different categories, and the Games, Entertainment, Tools, Social and Shopping are the categories with the highest number of apps having 217, 45, 28, 24 and 24 apps respectively.
    \item App Installation: Around 60\% of the apps have been installed more 10 million times.
    \item App Content Rating: Less than 40\% of the apps have content rating restrictions. Out of these, the majority have limited the audience to ``teens''.
    \item Chosen by editor: 106 apps are chosen by editor of the Google Play.
\end{enumerate}


\section{Results} \label{result}
In this section we present the results from our experiments to evaluate our research questions.

\subsection{Research Question 1} \label{result_rq1}
Regarding experiment configuration mentioned in Section~\ref{classifier}, the classifier labelled 10,972 reviews as the related ones to security and privacy concerns. This means in only around 0.5\% of all the reviews in our dataset, users write to mention a security/privacy concern. In order to provide more insights about the results, we have looked into the related reviews from different viewpoints:

\subsubsection{Apps’ categories}\label{rq1:cats} 
\begin{figure*}
    \centering
    \caption{Top categories based on the percentage of related reviews}
    \label{fig:categories}
    \begin{tikzpicture}
    \begin{axis}[
           symbolic x coords={Communication,Photography,Travel and Local,Social,House and Home,Dating,Productivity,Video Players,Business,Weather,Tools,Finance,News and Magazines,Personalization,Lifestyle,Music and Audio,Maps and Navigation,Shopping,Entertainment,Health and Fitness,Education,Auto and Vehicles,Sport,Art and Design,Books and Reference,Food and Drink,Libraries and Demo,Comics,Game,Beauty},
           legend image post style={mark=*},
           xtick pos=left,
           axis y line*=left,
           ybar,
           ymajorgrids,
           bar width=0.25cm,
           ymin=0, ymax=2.5,
           ytick={0, 0.5, 1, 1.5, 2, 2.5},
           yticklabels={0, 0.5, 1, 1.5, 2, 2.5},
           ylabel style={align=center},
           ylabel={Percentage of \\ Related Reviews},
           xtick=data,
           xticklabel style={
              inner sep=0pt,
              anchor=north east,
              rotate=45
              }
            ]
            \addplot[ybar,ybar legend,fill=blue!20] coordinates {(Communication,2.194601149)
                (Photography,1.491004274)
                (Travel and Local,1.4132226909999999)
                (Social,1.199956473)
                (House and Home,1.116071429)
                (Dating,0.8415147270000001)
                (Productivity,0.8145576659999999)
                (Video Players,0.806973983)
                (Business,0.7635002409999999)
                (Weather,0.751488095)
                (Tools,0.6847798)
                (Finance,0.648145946)
                (News and Magazines,0.553726429)
                (Personalization,0.522657189)
                (Lifestyle,0.504526135)
                (Music and Audio,0.450862314)
                (Maps and Navigation,0.37207917799999995)
                (Shopping,0.368615569)
                (Entertainment,0.360594088)
                (Health and Fitness,0.348782053)
                (Education,0.34598214299999996)
                (Auto and Vehicles,0.33994854799999996)
                (Sport,0.334821429)
                (Art and Design,0.290178571)
                (Books and Reference,0.272321429)
                (Food and Drink,0.246490194)
                (Libraries and Demo,0.242557883)
                (Comics,0.19243104600000002)
                (Game,0.17066473)
                (Beauty,0.166112957)
                };
            \end{axis}
\begin{axis}
    [axis y line*=right, xticklabels={},yticklabels={}]
  \end{axis}
        \end{tikzpicture}
\end{figure*}
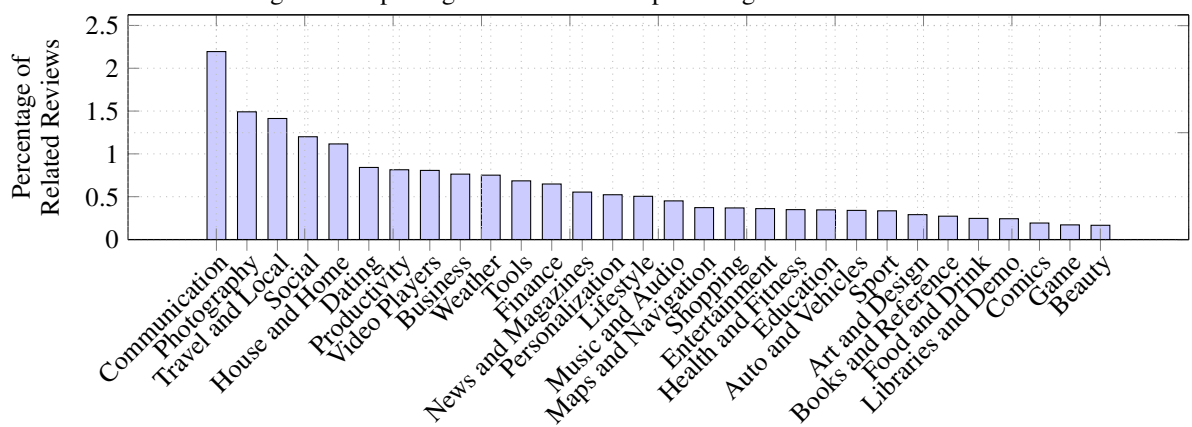

Figure~\ref{fig:categories} shows the percentage of related reviews per category. The top 30 categories are sorted and shown in this figure.

\subsubsection{Reviews’ Ratings}\label{rq1:ratings} 
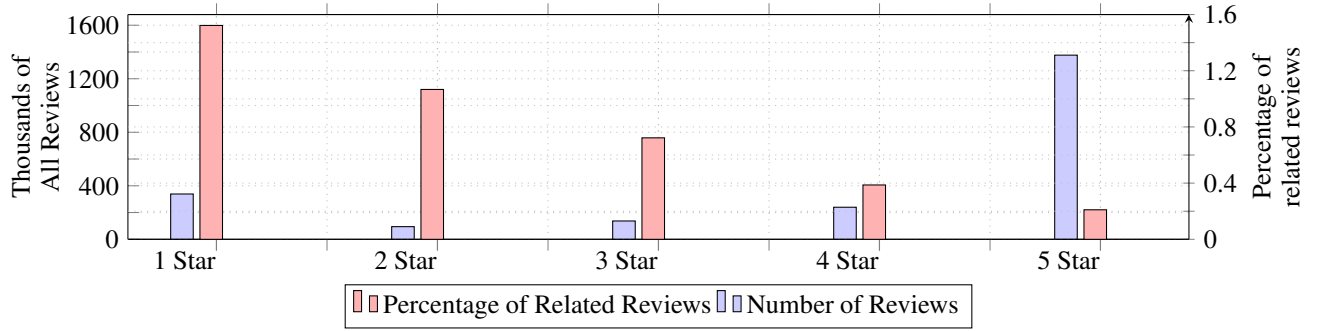
\begin{figure*}[]
    \centering
    \caption{Number of all the reviews and percentage of related reviews per rating stars}
    \label{fig:rating}
    \begin{tikzpicture}
    \begin{axis}[
          symbolic x coords={1 Star, 2 Star, 3 Star, 4 Star, 5 Star},
          legend style={at={(0.5,-0.1)},anchor=north , nodes={scale=0.5, transform shape}}, 
            legend image post style={mark=*},
              legend style={
              fill,
              at={(0.50,-0.2)},
              legend columns=2,
              legend cell align=left,
              anchor=north},
          xtick pos=left,
          axis y line*=left,
          ybar,
          bar shift = -13pt,
          ymajorgrids,
          bar width=0.3cm,
          ymin=0, ymax=1.6,
          ytick={0, 0.200000, 0.400000, 0.600000, 0.800000, 1.000000, 1.200000, 1.400000, 1.600000},
          yticklabels={0,  , 400,  , 800,  , 1200,  , 1600},
          ylabel style={align=center},
          ylabel={Thousands of \\ All Reviews},
          xtick=data,
          xticklabel style={
              inner sep=0pt,
              anchor=north east,
              rotate=0
              }
            ]
            \addplot[ybar,ybar legend,fill=blue!20] coordinates {(1 Star,0.338590)	(2 Star,0.094483)	(3 Star,0.136922)	(4 Star,0.239649)	(5 Star,1.376449)};\label{A}
    \end{axis}
    \begin{axis}[
              symbolic x coords={1 Star, 2 Star, 3 Star, 4 Star, 5 Star},
                legend style={
      fill,
      at={(0.50,-0.2)},
      legend columns=2,
      legend cell align=left,
      anchor=north
      },
              ybar,
              axis y line = right,
              ylabel style={align=center},
              ylabel={Percentage of \\ related reviews},
              bar shift = -2pt,
              bar width=0.30cm,
              ymin=0, ymax=1.6,
              ytick={0, 0.2, 0.4, 0.6, 0.8, 1,1.2, 1.4,1.6},
              yticklabels={0, , 0.4, , 0.8, ,1.2, ,1.6},
              xticklabels={},
  ]
    \addlegendentry{Percentage of Related Reviews}

  \addplot[ybar,ybar legend,fill=red!30] plot coordinates 
  {(1 Star,1.5224903280000002)	(2 Star,1.0668585879999999)	(3 Star,0.7223090520000001)	(4 Star,0.386815718)	(5 Star,0.21010585899999998)};
  \addlegendimage{/pgfplots/refstyle=A}
    \addlegendentry{Number of Reviews }
   \end{axis}
    \end{tikzpicture}
\end{figure*}

In our experiment we have classified the related reviews to security and privacy concerns, but we have not determined if the user is complaining of an app’s functionality regarding his concerns or is praising it. To provide a good insight, an efficient way can be leveraging the rating of reviews. We consider reviews with 4 and 5 stars as \emph{``positive''} reviews, reviews with 1 and 2 stars as \emph{``negative''} ones, and reviews with 3 stars as the \emph{``neutral''} reviews. Following this categorization, around 57\% of reviews tagged as negative, 9\% as neutral and 34\% as positive. Shown in Figure~\ref{fig:rating}, we have included the total number of reviews per rating in blue bars and the percentage of related reviews per rating stars with red bars.

\subsubsection{Apps’ Developers}\label{rq1:developers} 
\begin{figure*}
    \centering
    \caption{Top 20 developers based on number of apps along with the percentage of negative related reviews}
    \label{fig:developers}
    \begin{tikzpicture}
    \begin{axis}[
          symbolic x coords={Google LLC, VOODOO, Microsoft Corporation, Samsung Electronics, Amazon Mobile, Miniclip.com, Playgendary, Facebook, Lion Studios, Cheetah Games, Outfit7 Limited, Good Job Games, Azur Interactive Games, Playrix, Electronic Arts, Uber, Supercell, King, Amanotes, Ring.com},
           legend style={at={(0.5,-0.1)},anchor=top},
            legend image post style={mark=*},
          xtick pos=left,
          axis y line*=left,
          ybar,
          bar shift = 3pt,
          ymajorgrids,
          bar width=0.25cm,
          ymin=0, ymax=30,
           ytick={0, 5, 10, 15, 20, 25, 30},
           yticklabels={0, , 10, , 20, , 30},
          ylabel style={align=center},
          ylabel={Number of Apps},
          xtick=data,
          ytick align=inside,
          xticklabel style={
              inner sep=0pt,
              anchor=north east,
              rotate=60
              }
            ]
            \addplot[ybar,ybar legend,fill=blue!20] coordinates {(Google LLC,29) (VOODOO,13) (Microsoft Corporation,7) (Samsung Electronics,6) (Amazon Mobile,6) (Miniclip.com,5) (Playgendary,5) (Facebook,4) (Lion Studios,4) (Cheetah Games,4) (Outfit7 Limited,4) (Good Job Games,3) (Azur Interactive Games,3)	(Playrix,3) (Electronic Arts,3) (Uber,3) (Supercell,3) (King,3) (Amanotes,3) (Ring.com,2)};\label{A}
    \end{axis}
    \begin{axis}[
              symbolic x coords={Google LLC, VOODOO, Microsoft Corporation, Samsung Electronics, Amazon Mobile, Miniclip.com, Playgendary, Facebook, Lion Studios, Cheetah Games, Outfit7 Limited, Good Job Games, Azur Interactive Games, Playrix, Electronic Arts, Uber, Supercell, King, Amanotes, Ring.com},
              xticklabels={},
              ybar,
              axis y line = right,
              ylabel style={align=center},
              ylabel={Percentage of Negative\\ Related Reviews},
              bar shift = -4.6pt,
              bar width=0.25cm,
               ymin=0, ymax=2,
               ytick={0, 0.5, 1, 1.5, 2},
               yticklabels={0, 0.5, 1, 1.5, 2},
               ytick align=inside,
  ]
    \addlegendentry{Percentage of Negative Related Reviews}
  \addplot[ybar,ybar legend,fill=red!30] plot coordinates 
  {(Google LLC,0.470056572)	(VOODOO,0.321101705)	(Microsoft Corporation,0.461227063)	(Samsung Electronics,0.640345559)	(Amazon Mobile,0.193452381)	(Miniclip.com,0.084912406)	(Playgendary,0.111891868)	(Facebook,1.936422655)	(Lion Studios,0.17299107100000002)	(Cheetah Games,0.055803570999999996)	(Outfit7 Limited,0.050223214)	(Good Job Games,0.290178571)	(Azur Interactive Games,0.18818216)	(Playrix,0.11904761900000001)	(Electronic Arts,0.133928571)	(Uber,0.148820597)	(Supercell,0.081851328)	(King,0.044642857)	(Amanotes,0.052083332999999996)	(Ring.com,0.420792079)};
  \addlegendimage{/pgfplots/refstyle=A}
  \addlegendentry{Number of Apps}
   \end{axis}
    \end{tikzpicture}
\end{figure*}
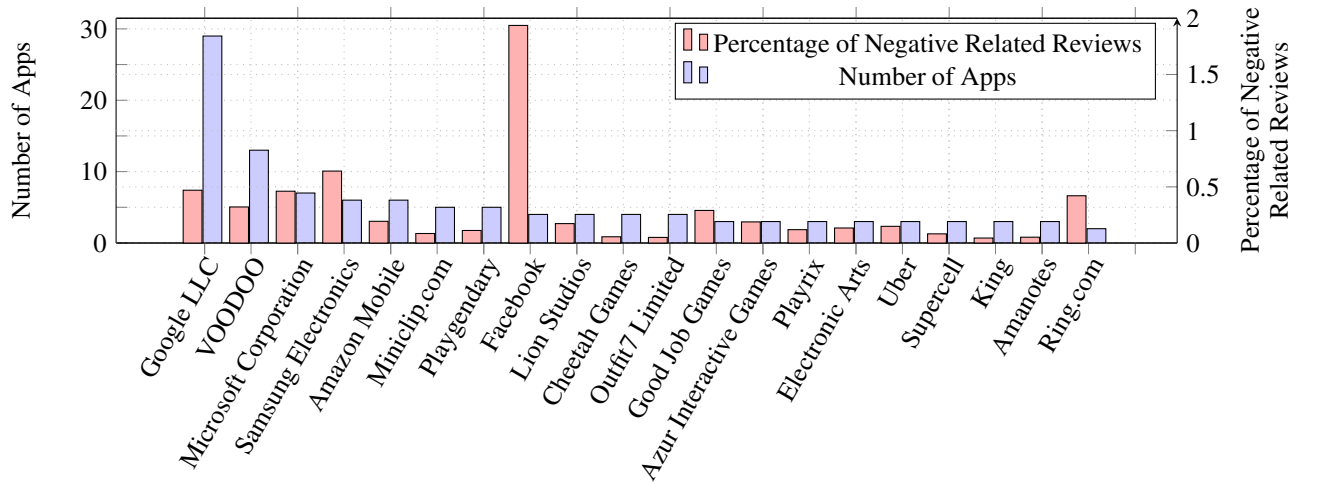

Another interesting perspective to notice is the role of developer in changing users’ thoughts. To assess this, Figure~\ref{fig:developers} shows the top 20 developers with highest number of apps along with the percentage of the negative related reviews for each.

\subsection{Research Question 2} \label{result_rq2}
To answer the second research question, we needed to assess the correlation between the users’ reviews and the functionality of the apps, determined from the analysis mentioned in Section~\ref{dynamic_analysis}.

At the very first step, we need to categorize apps based on users' thoughts about them. To achieve this, we tag related reviews in the same way mentioned for RQ1 in~\ref{rq1:ratings}. Then following the definitions mentioned in Table~\ref{tab:apps_tag_definition}, we tag the corresponding apps.

\begin{table*}[]
    \centering
    \normalsize
    \begin{center}
    \caption{Definition of the tags for apps}
    \label{tab:apps_tag_definition}
    \begin{tabular}{|c|p{9cm}|c|}
    \hline
    Tag name & \multicolumn{1}{c|}{Definition} & Number of apps \\ \hline
    Good & The ratio of positive reviews to all related reviews is greater than the same ratio for negative ones by 20\%. & 9 \\ \hline
    Neutral & The difference between the ratio of positive and negative reviews to all related reviews in less than 20\%. & 9 \\ \hline
    Bad & The ratio of negative reviews to all related reviews is greater than the same ratio for positive ones by 20\%. & 38 \\ \hline
    Not Discussed & The ratio of related reviews to all reviews is less than 1\%. & 483 \\ \hline
    \end{tabular}
    \end{center}
\end{table*}

Furthermore, from the dynamic analysis mentioned in Section~\ref{dynamic_analysis}, we found PII leaks over different hostnames for each app. We also found the number of \emph{dangerous} permissions asked by the apps. We call a permission ``dangerous'' when it needs two step confirmation from the user; in other words, it will be shown to the user at runtime through a pop-up dialogue. In order to make a meaningful comparison, we start considering different perspectives for categorizing apps using the analysis results. The description of perspectives along with some related statistics, the type of data, and their corresponding abbreviations (used for further comparison) are mentioned in Table~\ref{tab:perspectives_definition}.

\begin{table*}[]
\centering
\normalsize
\begin{center}
\caption{Description of the perspectives used for analysis}
\label{tab:perspectives_definition}
\setlength\tabcolsep{2.5pt}
\begin{tabular}{|p{6cm}|p{5cm}|p{3cm}|p{2cm}|}
\hline
\multicolumn{1}{c|}{Perspective} & \multicolumn{1}{c|}{Statistics} & Type of the data & Code \\ \hline
The number of hostnames contacted by an app (The chattiness the apps). & Average of 1.8 hostnames per app. & Numerical & CH \\ 
\hline
If an app is sending out the location information of the user or not. & 51 apps are sending. & Categorical & LOC  \\ \hline

If an app is sending out AAID along with another tracker, regardless of the hostname (Bridging AAID). & 100 apps are sending. & Categorical & BA \\ 
\hline
If an app is sending out AAID along with another tracker, over a single hostname (Bridging AAID). & 93 apps are sending. & Categorical & BAH \\ \hline
The maximum number of PIIs sent over a single hostname by an app. & Average of 0.7 PIIs per app. & Numerical & MPH  \\ \hline
The number of important permissions asked from user. & Average of 3.6 permissions per app. & Numerical & PE \\ 
\hline
\end{tabular}
\end{center}
\end{table*}

As the categorization of the apps based on reviews gives us categorical data, for correlation calculation we have categorical vs categorical data and categorical vs numerical data. As mentioned in Section~\ref{stats}, for the categorical vs categorical data we have used the symmetry approach of ``Cramer’s V'' (hereafter shown by CV) and asymmetry approach of ``Theil’s U'' (hereafter shown by TU1 and TU2\footnote{TU1 shows the correlation between first set and second one and TU2 shows the vice versa.}). For numerical vs categorical data also “Correlation Ratio” approach has been used (hereafter shown by CR).

The Table~\ref{tab:corr_result} shows the result. The third row shows the correlation between apps’ categories based on related reviews and categories of apps in different perspectives. Considering the Figure~\ref{fig:rating}, we noticed that the negative reviews form the majority of the related reviews. From this, we can assume that people do not usually praise apps for security and privacy concerns; therefore the apps with tag of ``Not discussed'' are actually ``Good''. Hence we replaced the tag of ``Not discussed'' to ``Good'' (in Table~\ref{tab:apps_tag_definition}). This provided us a new correlations whose results are listed in the last row. 

We have also used non-parametric statistical significance tests to assess the relation between categories of the apps in terms of reviews and in terms of their actual behavior. The Table~\ref{tab:stest_result} provides the p-values for both the \emph{MWU} and \emph{KWH} tests (row 3). Following the same assumption as above, in the last row we have the p-values for these tests, where all the apps with tag of ``Not discussed'' have been replaced with the tag of ``Good''.

\begin{table*}[]
\centering
\normalsize
\caption{Correlation achieved between app's categories based on reviews and apps' categories based on perspectives from functionality analysis}
\label{tab:corr_result}
\begin{center}
\setlength\tabcolsep{2.5pt}
\begin{tabular}{|c|c|c|c|c|c|c|c|c|c|c|c|c|}
\hline
Perspective & CH & \multicolumn{3}{c|}{LOC} & \multicolumn{3}{c|}{BA} & \multicolumn{3}{c|}{BAH} & MPH  & PE \\ \hline
Approach & CR & CV & TU1 & TU2 & CV & TU1 & TU2 & CV & TU1 & TU2 & CR & CR \\ \hline
Apps categorization based of reviews  & .080 & .132   & .018   & .024   & .025   & .010   & .009  & .000   & .008   & .007   & .055 & .365 \\ \hline
Apps categorization based of reviews* & .079 & .087   & .011   & .012   & .012   & .010   & .007  & .000   & .009   & .007   & .054 & .356 \\ \hline
\end{tabular}
\end{center}
\end{table*}

\begin{table*}[h]
\centering
\normalsize
\caption{P-values obtained from non-parametric statistical significance tests}
\label{tab:stest_result}
\begin{center}
\setlength\tabcolsep{2.5pt}
\begin{tabular}{|p{3cm}|c|c|c|c|c|c|c|c|c|c|c|c|}
\hline
\multicolumn{1}{c|}{Perspective} & \multicolumn{2}{c|}{CH} & \multicolumn{2}{c|}{LOC} & \multicolumn{2}{c|}{BA} & \multicolumn{2}{c|}{BAH} & \multicolumn{2}{c|}{MPH} & \multicolumn{2}{c}{PE} \\ \hline
\multicolumn{1}{c|}{Statistical Test} & KWH & MWU & KWH & MWU & KWH & MWU & KWH & MWU & KWH & MWU & KWH & MWU \\ \hline
Apps categorization based of reviews  & .706   & .359 & .041 & .022 & .330 & .170 & .647 & .331 & .467 & .238 & .649 & .331 \\ \hline
Apps categorization based of reviews* & .080   & .040  & .795 & .398 & .941 & .471 & .767 & .384 & .120 & .060 & .000 & .000 \\ \hline
\end{tabular}
\end{center}
\end{table*}

\section{Discussion} \label{sec_discussion}
After running the classifier, we found around 0.5\% of the all reviews are related to security and privacy concerns. We looked into the related reviews from different viewpoints. In terms of categories of the apps, although Game, Entertainment, Tools, Shopping, and Social are the categories with the highest reviews in our dataset, when it comes to the ratio of security/privacy related reviews, we see different categories on top of the list. It is not surprising to see categories such as Communication, Photography, Travel and Local, and Social are on top of the lists, as the users may be worried about the security of their personal information in communication and social mobile apps and also they may be concerned about the access of apps to some of their private information such as photos and location (cf.~Figure~\ref{fig:categories}). Considering the rating of the reviews in Figure~\ref{fig:rating}, it seems most of the time people complain about their concerns, as the number of related reviews with 1 star is the highest. In terms of the role of developer, we expected to see that the developers with higher number of apps in our dataset, have higher number of reviews and consequently the same ratio of negative related reviews. Initially we noticed that the top 20 developers with most apps in our dataset have the most reviews as well (with the same ordering). Then after plotting the percentage of negative related reviews, we noticed for some developers like ``Facebook'' and ``Samsung'', ratio of negative related reviews to all reviews are higher than expected (Figure~\ref{fig:developers}). This shows us that the developer of an app can potentially influence the users' perception.

After the process of dynamic analysis, we were able to match the actual behavior of the apps with users' judgment. The third row of the Table~\ref{tab:corr_result} shows the correlation between apps’ categories based on related reviews and categories of apps from the different perspectives. In the first look, we noticed that excluding the perspective of the last column (which relates to the number of permissions asked from users), in all the others, almost there is no correlation. Furthermore based on the results we obtained from the first research question, we assumed that people do not usually praise apps for security and privacy concerns and we changed the apps with tag of ``Not discussed'' to ``Good'' for better understanding. This resulted in new correlations shown in the last row. Again, excluding the last column, there is almost no correlation between different perspectives and apps’ categories based on users’ reviews. Considering that the only correlation happens based on the number of permissions asked from user, we can say the main criterion of the users for judging the functionality of the apps in terms of privacy and security is the number of permissions asked. In another viewpoint, users do not have any other promising criterion to help them decide how to feel about an app. For the significance tests (cf.~Table~\ref{tab:stest_result}), we assumed the significance level (the probability of rejecting the null hypothesis when it is true, denoted by $\alpha$) to be 0.05. We can see that the ``p'' value is very small for PE (row 4); so we can reject the null hypothesis and can confirm that the number of permissions asked is related to the categorization of the apps. Again for column LOC, the p value is smaller than $\alpha$ in row 3; hence, we can interpret that LOC also affects the app categorizations. This definitely makes sense because if an app sends out location details, it needs to ask for permission from the user in most of the cases. Therefore, the results under LOC are affected in the same way as PE. So, in these aspects, we can say that the users' reviews can be a signal of the app's behavior. On the other side, if we consider the p-values under CH, it is greater than $\alpha$ for the majority of the cases; hence we failed to reject the null hypothesis. We have the same results for BA, BAH and MPH perspectives. So in these cases, we can say that the users' review are not a good indicator for the behavior of the apps.

The main limit of our study was the restriction of Google API for accessing to reviews of an app. For some of the famous apps, we only had access to the latest reviews for last few months. The reviews in short time intervals can be biased and may not provide a fair representation of users' perception.

We found that the number of permissions can considerably influence users' thoughts. Therefore, a main step for future can be evaluating if the higher number of permissions necessarily results in worse functionality in terms of security and privacy. Along with this, investigating the most efficient practices for addressing privacy/security concerns while getting high number of permissions, is always an interesting topic. All these ideas can be equipped by approaches for making users to feel safe and comfortable while granting permissions. On the other hand, developing tools and guidelines to increase the number of users' criterion for judging their privacy violations can be a good step in the future. 

\section{Conclusion} \label{sec_conclusion}

In our study, we initially scraped details and reviews of 539 top free mobile applications of the Google Play Android app market. We made our dataset with the total number of 2,186,093 of reviews. After pre-processing phase, we ran our classifier, SVM, which provided us the best performance according to 10-fold cross validation. In our experiment, 0.5\% of reviews are classified as related to security and privacy concerns. Our analysis showed that users are more concerned for the apps from categories such as communication, photography, travel, and social. They also mostly provide their security/privacy related reviews along with low ratings (1 and 2 stars). We also noticed that users may get influenced by the developer of an app. It means, they may trust or distrust a developer.

In order to assess the judgment of users, we performed dynamic analysis to see the actual behavior of the apps. We found a correlation between the judgment of users for an app and the number of permissions asked from user by it. We also noticed some correlation between the categories of apps based on user reviews to those apps that send out location information. In all other perspectives such as ``number of domains contacted'' and ``bridging AAID'', no significant correlation was observed. This can be originated from the fact that permissions are the main criterion for judging the amount of access of the apps to personal data.


\bibliography{MobileSecurityArXiv.bib}

\vspace{12pt}

\bibliographystyle{unsrt}  
\end{document}